# Lightning-triggered electroporation and electrofusion as possible contributors to natural HGT among prokaryotes

Tadej Kotnik

*University of Ljubljana, Faculty of Electrical Engineering, Tržaška 25, SI-1000 Ljubljana, Slovenia*

**Phylogenetic studies show that horizontal gene transfer (HGT) is a significant contributor to genetic variability of prokaryotes, and was perhaps even more abundant during the early evolution. Hitherto, research of natural HGT has mainly focused on three mechanisms of DNA transfer: conjugation, natural competence, and viral transduction. This paper discusses the feasibility of a fourth such mechanism — cell electroporation and/or electrofusion triggered by atmospheric electrostatic discharges (lightnings). A description of electroporation as a phenomenon is followed by a review of experimental evidence that electroporation of prokaryotes in aqueous environments can result in release of non-denatured DNA, as well as uptake of DNA from the surroundings and transformation. Similarly, a description of electrofusion is followed by a review of experiments showing that prokaryotes devoid of cell wall can electrofuse into hybrids expressing the genes of their both precursors. Under sufficiently fine-tuned conditions, electroporation and electrofusion are efficient tools for artificial transformation and hybridization, respectively, but the quantitative analysis developed here shows that conditions for electroporation-based DNA release, DNA uptake and transformation, as well as for electrofusion are also present in many natural aqueous environments exposed to lightnings. Electroporation is thus a plausible contributor to natural HGT among prokaryotes, and could have been particularly important during the early evolution, when the other mechanisms might have been scarcer or nonexistent. In modern prokaryotes, natural absence of the cell wall is rare, but it is reasonable to assume that the wall has formed during a certain stage of evolution, and at least prior to this, electrofusion could also have contributed to natural HGT. The concluding section outlines several guidelines for assessment of the feasibility of lightning-triggered HGT.**

## Evolution and Gene Transfer

**The Phylogenetic Tree is Not Exactly a Tree.** During the last century and a half, the concept of biological evolution has traversed a path from a widely contested conjecture to a broadly accepted scientific theory. In addition to an increasing body of paleontological evidence, during the last two decades this concept has also been backed by direct experimental proof (1). During this same period, however, another concept that seemed to be a logical corollary of evolution — that there is a universal phylogenetic tree leading from the last common ancestor, purely through branching, to all other organisms to have lived on Earth ever since — has started to unravel. Nucleic acid sequencing, which was central in determination of genetic relatedness among organisms, including identification of archaea as a separate taxonomic domain (2,3), started also to provide incontrovertible evidence that the phylogenetic structure being charted is not exactly a tree, as the branches can interconnect laterally and even fuse pairwise (Fig. 1).

**The Phylogenetic Tree is Rather a Phylogenetic Network.** Strictly speaking, horizontal gene transfer (HGT) among bacteria, first elucidated in the early 1960s (4), already implied lateral interconnections between branches within the bacterial domain, while recognition of eukaryotic mitochondria and plastids as endosymbiotic descendants of α-proteobacteria (5,6) and cyanobacteria (7,8), respectively, implied two lateral fusions between major branches of bacterial and eukaryotic domains. Still, these were generally viewed as exceptional events, and conjectures that HGT could have played a prominent role in the evolution (9,10) were rare. As technology reached a stage allowing for extensive DNA sequencing, however, it emerged that a number of genes in eukaryotes are absent from archaeal genomes, yet present in phylogenetically more distant bacteria (11,12), and it also became increasingly clear that the phylogenetic trees charted from different genes can differ considerably (13,14). These findings implied rather potently that HGT is a significant contributor to genetic variability of unicellular organisms, and suggested that it was perhaps even more abundant in the evolution of early-life prokaryotes (15–17).

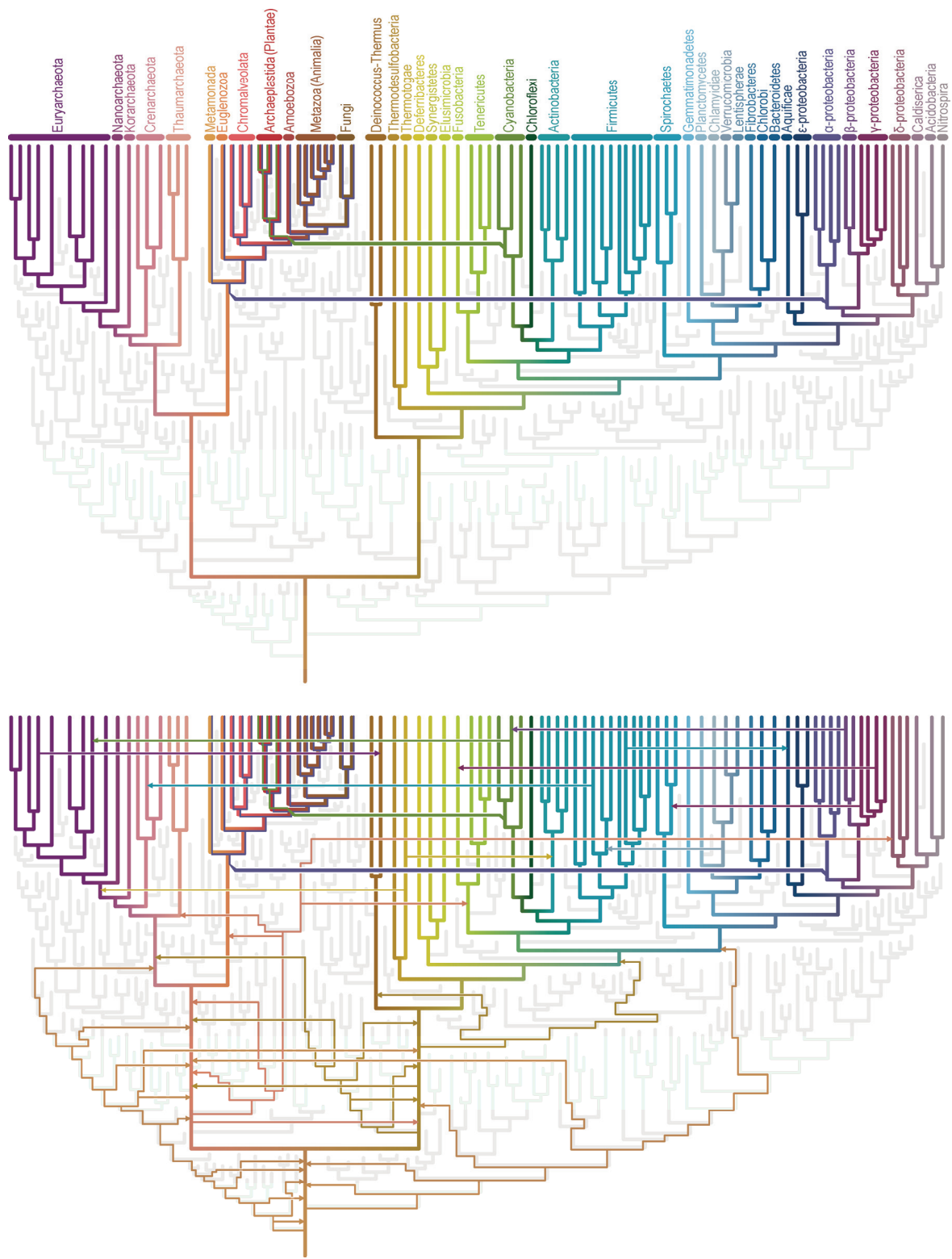

**Fig. 1.** A sketch of a current view of the phylogenetic "tree", with the vertical axis corresponding to time. Top: *The Phylogenetic Tree of Cell Divisions and Speciations* for the major surviving (colored) and extinct (light gray — entirely hypothetical) lineages, disregarding all HGT except for mergers of entire genomes (shown are the incorporations of mitochondrial and plastid ancestors into eukaryotes). Phylogenetic proximity of several taxa shown is still far from the consensus, and the future contemporary views will likely differ from this one noticeably. The extinct branches will, however, remain largely unknown, as the vast majority of extinct organisms' genomes have by now decomposed completely. Bottom: *The Phylogenetic Network* obtained by including HGT, but with only its tiny fraction displayed to avoid unintelligibility. The very local topological structure of the branches (particularly within the eukaryotic domain) may still be that of a tree, but there are numerous transfers of genome fragments, particularly between prokaryotes. HGT from the now extinct lineages to the surviving ones (the transfers shown are, again, entirely hypothetical) also contributed to their genomes. Inspired by graphics of Martin (18), Ford Doolittle (19), Zhaxybayeva and Gogarten (14), Ciccarelli et al. (20); and accounting for some results from recent investigations of archaea (21,22), bacteria (23–26), and HGT among them (27).

**Three Natural Mechanisms of HGT.** Hitherto, research of natural HGT has mainly focused on three mechanisms of DNA transfer: bacterial conjugation, natural bacterial competence for DNA uptake, and viral transduction. Conjugation involves DNA relaxases and P-type or F-type pilin filaments in the donor bacterium (28,29), competence is based on DNA translocases and type-IV pilin filaments in the acceptor bacterium (29,30), while transduction requires a bacteriophage to attach its base plate to the bacterium, penetrate the bacterial wall and membrane with its tail, and inject its genetic material. Each of these rather intricate mechanisms must have developed during a certain stage of evolution, and whether — and how — HGT could take place prior to these stages is an open question.

**Is There a Fourth Natural Mechanism?** This paper investigates the possibility of a fourth mechanism of natural HGT — membrane electroporation, and possibly also cell electrofusion, triggered by atmospheric electrostatic discharges (lightnings). A possible role of electroporation in bacterial evolution has been contemplated by several researchers (31–33), but no comprehensive investigation of this topic seems to have been published to date. This paper aspires to provide at least a starting point for such a scientific venture. Electroporation is first described as a phenomenon and then as a method for DNA release and/or uptake and transformation. Similarly, electrofusion is outlined as a phenomenon and then as a method for production of cell hybrids. Then, feasibility of electroporation and electrofusion as natural mechanisms of HGT triggered by lightnings is considered. In conclusion, the paper outlines several guidelines and possible setups for experimental verification of the presented hypothesis.

## Electroporation and Gene Transfer

**Membrane Electroporation as a Phenomenon.** An exposure of biological membranes to a sufficiently high electric field leads to a rapid and large increase of their electric conductivity and permeability. This effect — referred to as membrane electroporation — can be either reversible or irreversible, and was first reported for excitable cells in 1958 (34), for nonexcitable cells in 1967 (35), for lipid vesicles in 1972 (36), and for planar lipid bilayers in 1979 (37).

Both theoretical considerations (38) and molecular dynamics simulations (39–43) imply that on the molecular scale, electroporation is the result of metastable aqueous pores formed by penetration of water molecules into the lipid bilayer and by subsequent reorientation of the adjacent lipids with their polar headgroups towards these water molecules. A molecular-level scheme of this stochastic process and an example of its atomic-level molecular dynamics simulation are shown in Fig. 2.

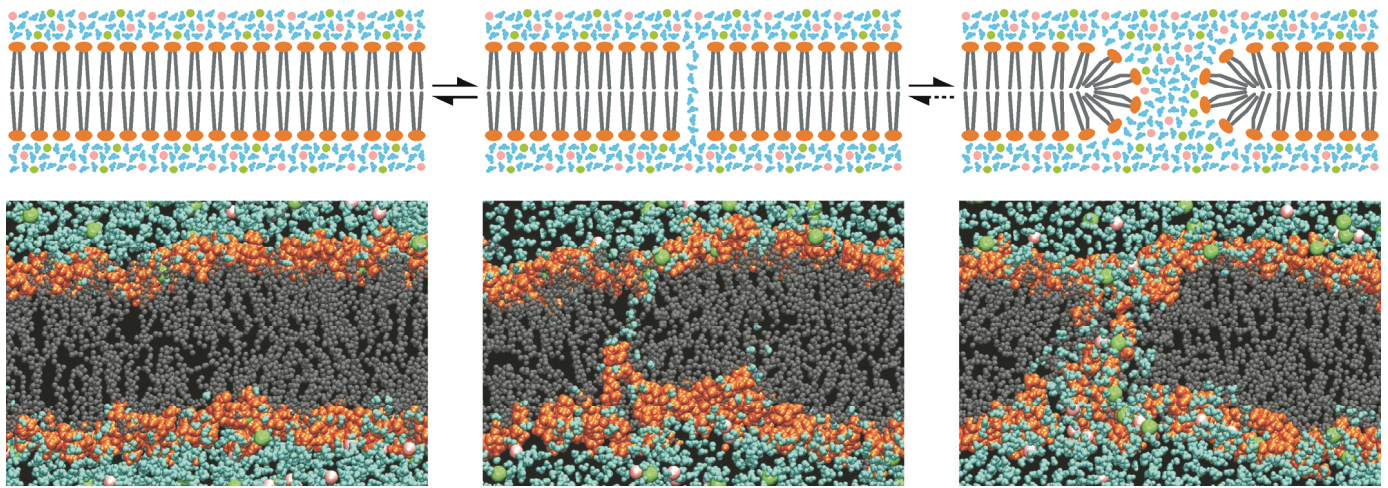

**Fig. 2.** An idealized molecular-level scheme (top) and an atomic-level molecular dynamics simulation (bottom) of electroporation, with the electric field perpendicular to the bilayer plane, and with the bilayer surrounded by a saline solution. In the simulation, a POPC bilayer is exposed to a field of 4 MV/cm, and the snapshots correspond to 0, 0.15, and 0.50 ns after the field is turned on. Left: the intact bilayer. Middle: water molecules start penetrating the bilayer, forming a water wire. Right: the lipids adjacent to the water wire start reorienting towards them with their polar headgroups, stabilizing the aqueous pore and allowing further water, as well as other polar molecules and ions to enter the pore. The atoms of the lipid headgroups and tails are shown in orange and gray, respectively, water molecules in cyan, sodium ions in green, and chloride ions in pink. The bottom panel is reprinted from Kotnik et al. (43) with permission.

For *external* electric fields of several kV/cm (which on the plasma membrane of a cell several µm large induce *transmembrane* electric fields of several MV/cm, and corresponding transmembrane voltages from several hundred mV up to several V), formation of metastable aqueous pores is completed within nanoseconds after the onset of the field (39–43). The pores thus formed in the cell plasma membrane provide a pathway for transport of a wide range of molecules, including DNA, into (44) and out of the cell (45). Electroporation is a physical phenomenon, and can as such occur in the lipid bilayer of the membranes of all prokaryotic and eukaryotic cells. Pore formation is governed by statistical thermodynamics (38,46), so it is not strictly a threshold event, in the sense that the pores could only form in electric fields exceeding a certain value. Nonetheless, electroporation-mediated transport across the membrane is strongly correlated with the transmembrane voltage induced by the external electric field (47), which is in turn proportional to this field (47–51). There are four general ranges of electric field, each characterized by typical properties of the pores being formed and/or the related molecular transport (Fig. 3):

- In the low range, the pores — even if formed — are too small and too short-lived for molecular transport across the membrane to proceed through them.
- In the intermediate range, most of the cells exposed are electroporated reversibly; the pores provide a temporary pathway for molecular transport, but after the end of the electric pulse, they gradually reseal and the transport ceases.
- In the high range, the cells are electroporated irreversibly; the pores do not reseal, but instead keep expanding, resulting in disintegration of the cells and release of intracellular contents; still, there is generally no thermal damage to the released molecules.
- In even stronger fields, the electric currents cause a temperature increase sufficiently high for thermal damage to the released molecules, including DNA denaturation (melting).

The bounds of these ranges depend on the duration of the exposure to the electric field, on the type of the cells exposed, and on the properties of the medium in which the exposure takes place. Moreover, the cells within the exposed population generally vary in size, as well as in their orientation with respect to the field direction (unless they are purely spherical). Together with the stochasticity of pore formation, this results in overlapping bounds of the regions described above: in a certain range of electric fields, some cells are porated and others not; in another range, some are porated irreversibly and others reversibly. Moreover, the longer the pulses, the lower the minimal field at which thermal damage occurs. Still, some general limits can be outlined. Thus if prokaryotes are exposed to submillisecond electric pulses (this includes lightnings), for pulse amplitudes up to hundreds of V/cm electroporation is undetectable; within a subrange of the range spanning from hundreds of V/cm up to tens of kV/cm it is mostly reversible; and above this subrange it becomes mostly irreversible, with thermal damage only occurring at hundreds of kV/cm (Fig. 3).

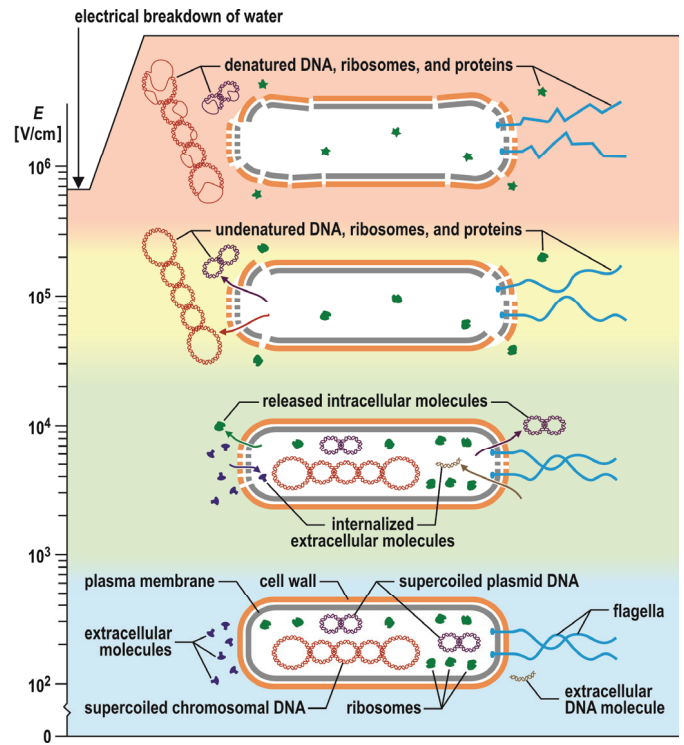

**Fig. 3.** Electroporation-mediated molecular transport as a function of the external electric field (oriented here horizontally) to which the cell is exposed. Reversible poration allows for uptake of extracellular DNA, as well as possible release of plasmid DNA. Irreversible poration results in unrestricted release of intracellular contents, including chromosomal DNA. With even stronger fields, temperature increase leads to thermal damage, including DNA denaturation, and finally to electrical breakdown (ionization) of water.

Since its discovery, reversible membrane electroporation has been steadily gaining ground in various areas of biology, biotechnology, and medicine, becoming an established method for introduction of chemotherapeutics into tumor cells (52) and a promising technique for gene therapy devoid of the risks caused by viral vectors (53).

**Irreversible Electroporation as an Initiator of DNA Release from Prokaryotes.** Perhaps the earliest reported use of high-voltage electric pulses for killing of microorganisms dates back to 1896, when the Louisville Water Company studied various methods of purifying river water (54). The first scientific study of destruction of bacteria by irreversible membrane electroporation was published in 1967, showing that the lethal effect is nonthermal and due to extensive membrane disruption that results in the leakage of intracellular contents, including DNA (35,45). Since then, irreversible electroporation has become a well-known method for nonthermal deactivation of microorganisms, particularly in food preservation (55), and it is also an efficient tool for extraction of biomolecules (56).

**Reversible Electroporation as an Efficient Technique for Transformation of Prokaryotes.** Artificial electroporation-induced uptake of DNA with subsequent expression — gene electrotransfer — was first achieved in mammalian cells in the early 1980s (45,57,58), while in bacteria the first studies suggested this to be achievable only after a complete removal of the cell wall (59). The early inability to obtain gene electrotransfer in bacteria with a cell wall was, however, largely due to insufficient available electric fields, and the development of stronger pulse generators quickly led to transformation of intact *Lactococcus lactis*, *Escherichia coli*, *Lactobacillus casei*, and *Streptococcus thermophilus* in 1985–1987, with more than thirty bacterial species following in 1988 (60), and the first archaeon (*Methanococcus voltae*) in 1991 (61).

Gene electrotransfer is effective both in bacteria with a Gram-negative and in those with a Gram-positive cell wall, although the latter generally require higher fields and/or yield lower efficiencies as measured by the number of transformed colony-forming units per µg of DNA (typically up to $10^8$–$10^{10}$ CFU/µg DNA for Gram-negative, and up to $10^6$–$10^7$ CFU/µg DNA for Gram-positive bacteria, when electrotransformed with plasmid DNA), which is attributed to the thicker peptidoglycan layer of the Gram-positive wall (60,62). In the same vein, bacteria possessing an outer polysaccharide capsule are generally electrotransformed with even lower efficiencies, but these can still exceed $10^4$ CFU/µg DNA (again, with plasmid DNA) for bacteria in the exponential growth phase (63), in which the capsular synthesis rate decreases (64,65).

Efficiency of gene electrotransfer in bacteria also depends strongly on the molecular form of DNA being transferred. Generally, the efficiency is the highest for supercoiled circular double-stranded DNA (the indigenous form of plasmid and chromosomal DNA in many prokaryotes), somewhat lower for relaxed circular double-stranded DNA, much lower for circular single-stranded DNA (indigenous to most ssDNA viruses) and linear double-stranded DNA with homologous ends (indigenous to eukaryotes), and lower still for linear double-stranded DNA with nonhomologous ends. Thus in a comparative study performed on *E. coli*, efficiency of transformation with the pACYC177 plasmid dsDNA was $1.9 \times 10^9$ CFU/µg DNA for the supercoiled circular form, $9.6 \times 10^8$ CFU/µg DNA for the relaxed circular form, and $1.2 \times 10^5$ CFU/µg DNA for the linear form with homologous ends (66). In the same study, efficiency of transfection of *E. coli* with the circular ssDNA of α3 and φKh-1 phages was $8.0 \times 10^6$ and $3.2 \times 10^5$ CFU/µg DNA, respectively (66), and in another study on *E. coli*, the efficiencies of transformation with two different linear forms of the pBR322 plasmid dsDNA with nonhomologous ends were $1.6 \times 10^3$ and $4.0 \times 10^2$ CFU/µg DNA (67). It should be noted that with such a wide range of transformation efficiencies, techniques adequate for assessing the higher end of this range generally leave the results in the lower end undetected. This could have contributed to some assertions that gene electrotransfer in bacteria is unachievable with linear dsDNA or even chromosomal circular dsDNA (68), and that bacteria without an indigenous plasmid cannot be transformed even with plasmid circular dsDNA (69), although it is impossible to disprove rigorously that for some bacteria this may also be the actual case.

For DNA concentrations spanning from pg/ml up to µg/ml, efficiency of gene electrotransfer in bacteria is roughly constant, implying that within this range and under fixed experimental conditions, for each bacterium the probability of being transformed increases roughly proportionally to the concentration of DNA surrounding it (70). The optimal parameters of the electric pulses used to achieve gene electrotransfer vary with bacterial species and even strain, but generally, pulse amplitudes (electric fields) range from 2 to 30 kV/cm, and pulse durations from milliseconds to tens of milliseconds (60,62,71). In eukaryotic cells, the efficiency of gene electrotransfer, particularly at low DNA concentrations, is improved if the electroporating pulse is followed by a contiguous, longer but much weaker pulse (tens or hundreds of milliseconds, tens of V/cm) that exerts an electrophoretic drag on the DNA molecules (72,73). This effect does not seem to have yet been investigated in bacteria or archaea, but it may still be worth noting that with the now prevailing rectangular-pulse generators, the electrophoretic "tail" of the pulse has to be formed by appending a second pulse, while with the simpler exponential-decay-pulse generators it was inherent to the pulse shape, which is also the case for natural discharges, including lightnings.

Efficiency of gene electrotransfer in bacteria can be improved by hyperosmolarity of the medium, typically achieved by dissolving sorbitol or mannitol at 0.5–1.5 M concentrations (74,75). In nature, hyperosmolarity of aqueous media is generally a consequence of high concentrations of salts, but the efficiency of gene electrotransfer in such hyperosmolar media does not seem to have yet been studied, likely due to the fact that for lab transformation protocols, similarity to natural conditions is less important than efficiency and practical feasibility. With respect to the latter, unlike sorbitol or mannitol, added salts can increase the electric conductivity of the medium considerably; for a fixed electric field (as delivered by voltage pulse generators), this increases the electric current and the heating of the medium, while for a fixed electric current (as delivered by current generators), this reduces the electric field induced by this current.

As described above, the highest efficiencies of gene electrotransfer in bacteria are generally achieved with plasmid DNA. Among strains of the same species, transfer can be efficient also with unaltered indigenous plasmids, but between distant species, and particularly between phyla, the highest efficiency is usually obtained with artificially engineered chimeric plasmids (shuttle vectors). The principal aim of implementing gene electrotransfer in bacteria is generally to achieve the most efficient transformation possible; furthermore, in many applications the transferred gene originates from a eukaryote, which requires the plasmid to be altered at least by incorporation of this gene. As a consequence, in the large majority of existing reports on successful gene electrotransfer in prokaryotes, the main result was obtained with engineered DNA, and there has not been much motive for a systematic investigation of feasibility of interspecies electrotransfer of unaltered natural DNA. Still, as Table I shows, gene electrotransfer can in general occur in species of many archaeal and bacterial phyla.

Table I. A sample of reports on successful gene electrotransfer into species of various prokaryotic phyla.

| phylum | species |
|---|---|
| **Archaea** | |
| Crenarchaeota | *Sulfolobus solfataricus* (76), *Sulfolobus acidocaldarius* (77) |
| Euryarchaeota | *Methanococcus voltae* (61), *Pyrococcus furiosus* (78) |
| **Bacteria** | |
| Actinobacteria | *Mycobacterium smegmatis* (79), *Corynebacterium diphtheriae* (80), *Brevibacterium lactofermentum* (81), … |
| Bacteroidetes | *Bacteroides uniformis* (82), *Bacteroides fragilis* (83), *Prevotella ruminicola* (84), … |
| Chlamydiae | *Chlamydia trachomatis* (85), *Chlamydia psittaci* (86) |
| Chlorobi | *Chlorobium vibrioforme* (87) |
| Cyanobacteria | *Fremyella diplosiphon* (88), *Synechococcus elongatus* (89) |
| Deinococcus-Thermus | *Deinococcus geothermalis* (90), *Thermus thermophilus* (91) |
| Firmicutes | *Lactobacillus casei* (92), *Enterococcus faecalis* (93), *Bacillus cereus* (94), *Streptococcus pyogenes* (95), *Clostridium perfringens* (96), … |
| Fusobacteria | *Fusobacterium nucleatum* (97) |
| Planctomycetes | *Planctomyces limnophilus* (98) |
| Proteobacteria | *Escherichia coli* (99), *Campylobacter jejuni* (100), *Sinorhizobium meliloti* (101), *Salmonella enterica* (102), *Yersinia pestis* (103), … |
| Spirochaetes | *Borrelia burgdorferi* (104), *Serpulina hyodysenteriae* (105) |
| Tenericutes | *Mycoplasma pneumoniae* (106) |
| Thermotogae | *Thermotoga maritima* (107) |

# Electrofusion and Gene Transfer

**Cell Electrofusion as a Phenomenon.** An exposure of biological cells to electric pulses with amplitude and duration sufficient to induce membrane electroporation has been shown to also cause the exposed cells to fuse with each other. This effect — cell electrofusion — was first demonstrated for plant protoplasts in 1978, for anucleate animal cells (erythrocytes) in 1980, for lipid vesicles in 1981, and for nucleated animal cells (fibroblasts) in 1982 (108–111).

Experiments show that electrofusion of two cells can occur if they are in direct contact during the exposure to the electric pulses, but also if they are brought into direct contact within seconds or even minutes after the exposure (112,113). In the former case, pores in the contact area of the two lipid bilayers will often form in coaxially aligned locations, as formation of a pore in one bilayer doubles the voltage induced across the other bilayer in its region adjacent to the pore; such double-membrane pores connect the interiors of the two cells, and it was proposed that lasting electrofusion could result from their coalescence (114). Still, thermodynamic considerations infer that this form of electrofusion is energetically very demanding and hence rare; together with the ability of bilayers to fuse also when brought into contact after pulse delivery (when close coaxial alignment of pores is much less likely), this implies that double-membrane-pore coalescence is not the only — and unlikely to even be the main — underlying mechanism of electrofusion (112,115).

Experiments also show that in electrofusion of two lipid bilayers, the monolayers in direct contact often fuse first, while the other two monolayers still appear intact (116). This suggests that on the molecular scale, electrofusion proceeds in the same three stages that are now broadly recognized in the physiological processes of membrane fusion: (i) the two monolayers in direct contact, of which at least one must be locally destabilized, fuse within a small area containing the local instability, forming a stalk; (ii) the fused monolayers start moving apart radially, bringing the other two monolayers into contact, thereby forming a disk-shaped diaphragm; (iii) the diaphragm ruptures, thus forming a pore that connects the interiors of the two cells (117–119). A molecular-level scheme of the fusion process is shown in Fig. 4. The principal difference between physiological fusion and electrofusion thus lies in the catalyst of local monolayer destabilization that initiates the first stage of fusion — in the former case various fusogenic membrane proteins (119–121), and in the latter the electric pulses (115).

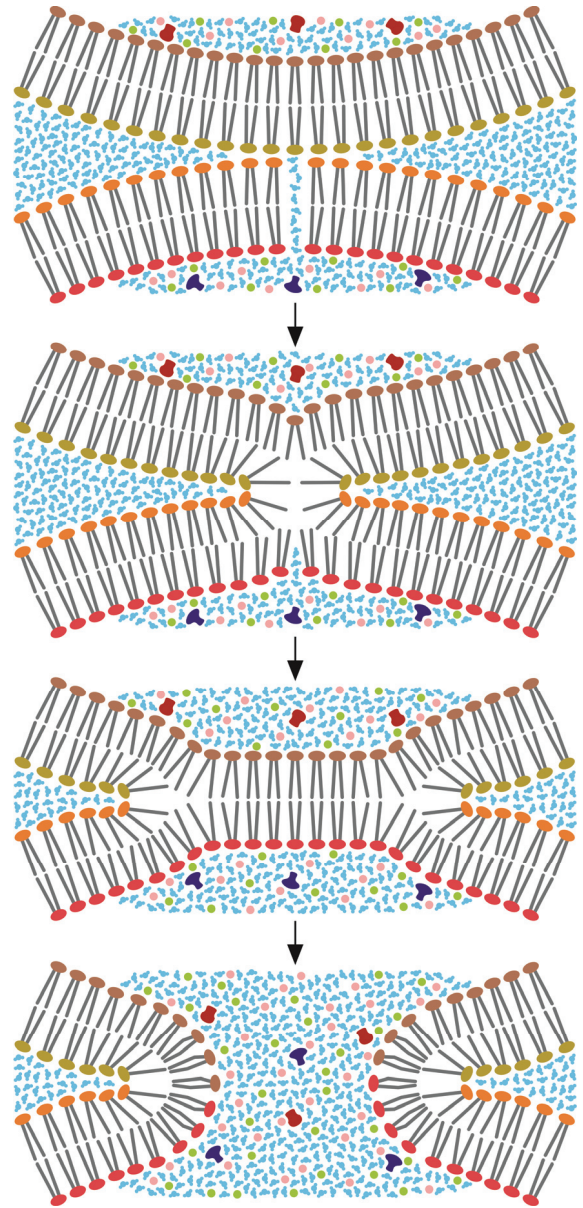

**Fig. 4.** An idealized molecular-level scheme of cell fusion initiated by electroporation. From top to bottom: two membranes, one electroporated (shown is a water wire as the most unstable structure in the process of pore formation and resealing, see Fig. 2), come into contact; the proximal monolayers fuse locally, forming a hemifusion stalk; the fused monolayers move apart radially, and the distal monolayers come into contact, forming a disk-shaped hemifusion diaphragm; the diaphragm ruptures, forming a fusion pore and allowing the contents enveloped by the two membranes to mix.

As with electroporation, molecular dynamics simulations are an important tool for verifying the current view and improving the understanding of molecular-scale events underlying electrofusion and fusion of membranes in general. Compared to a simulation of a pore forming in a bilayer, a similarly accurate and consistent simulation of two bilayers fusing requires a much larger number of lipid molecules, and generally also longer time scales — particularly if also the formation of the initial local monolayer instability, e.g. by electroporation, is to be simulated — and therefore a larger number of timeframes. As a consequence, the details achievable today in atomic-level molecular dynamics studies of electroporation may still be years away for electrofusion. Still, the results obtained to date both with coarse-grained (122,123) and atomic-level simulations (124,125) generally confirm the description of fusion as a three-stage, stalk-diaphragm-pore process outlined in Fig. 4.

Combined with suitable methods of bringing the cells into close contact, particularly by means of dielectrophoresis (126) or in microfluidic chambers (127,128), electrofusion has become an efficient method for cell hybridoma production (129–131), serving as an alternative or a complement to the techniques employing viral fusogenic proteins and/or polyethylene glycol.

**Electrofusion as a Technique for Hybridization of Prokaryotic Protoplasts and Spheroplasts.** Similarly to electroporation of bacteria, also electrofusion of bacteria was first reported to be achievable only after a removal of the cell wall — i.e., either in protoplasts, in which the wall is removed entirely, or in spheroplasts, in which the wall is removed partially (to the extent that they become roughly spherical in shape). Thus, in 1983 protoplasts of two strains of *Bacillus thuringiensis* were successfully fused into viable hybrids that expressed genes of their both precursors (132). In the years that followed, this was also achieved with spheroplasts of *Escherichia coli* and *Salmonella typhimurium* (133), protoplasts of two strains of *Corynebacterium glutamicum* (134), protoplasts of *Lactobacillus acidophilus* and *Streptococcus lactis* (135), and protoplasts of two strains of *Lactobacillus plantarum* (136). However, unlike electroporation, electrofusion of bacteria with an intact wall can not be achieved by increasing the electric field, even up to the levels at which all the exposed bacteria lose their viability (135); the wall thus rather appears to be an essential impediment to fusion of lipid bilayers, as it precludes a direct contact between them (see Fig. 4). In all the discussed studies, the optimal pulse amplitudes and durations for electrofusion of bacterial protoplasts and spheroplasts were in the range from 3 to 12 kV/cm and from 15 to 150 µs, respectively, in which bacteria are typically electroporated reversibly (see Fig. 3).

## Electroporation, Electrofusion, and Evolution

**Electroporation and Electrofusion as Possible Natural Mechanisms of HGT.** In an aqueous environment struck by an atmospheric electrostatic discharge (lightning), downward and outward from the point of entry of the discharge, the electric field generated by it is roughly inversely proportional to the square of the distance from this point. The electric field thus decreases radially in a continuous and monotonic manner, so there is generally a region where this field is insufficient for substantial heating, yet sufficient for irreversible membrane electroporation, and adjacent to this region there is another where the field is insufficient for irreversible, yet sufficient for reversible poration. Hence the DNA released from electroporated organisms in these two regions is not subject to thermal denaturation, and the electroporated organisms in the latter region are generally capable of DNA uptake and expression (Fig. 5, left). Similarly, two prokaryotic organisms naturally lacking a cell wall, in (at least) one of which the membrane is reversibly electroporated, can electrofuse into a hybrid organism expressing the genetic material of their both precursors (Fig. 5, right).

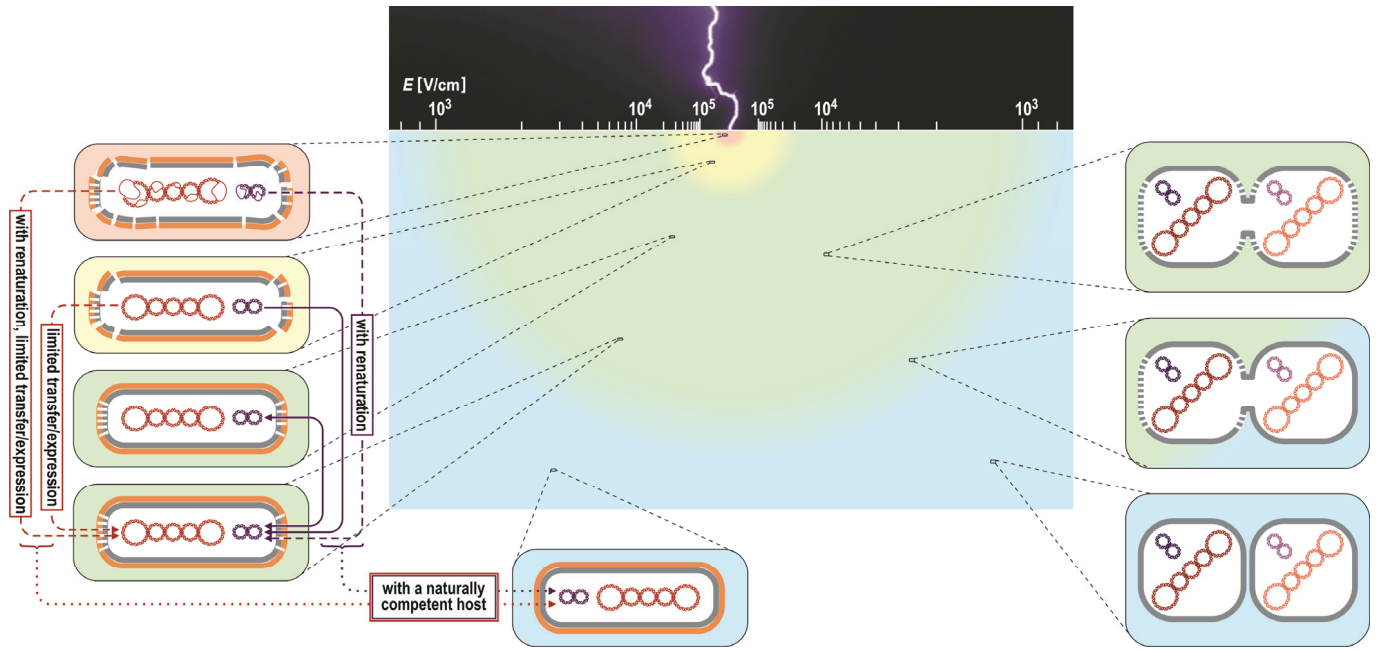

**Fig. 5.** Electroporation (left) and electrofusion (right) as potential natural mechanisms of HGT. DNA is released from organisms in the regions of thermal damage (pink) and nonthermal irreversible electroporation (yellow), and plasmid DNA also in the region of reversible electroporation (green). For host organisms without natural competence, transformation is limited to the region of reversible electroporation, while naturally competent hosts can also be transformed by the released DNA in the region without electroporation (blue). For successful electrofusion, the fusing organisms must be devoid of the cell wall, and at least one of them has to be electroporated.

**Electric Fields And Temperature Increases Caused by Lightnings.** Atmospheric electrostatic discharges are classified according to the polarity of the net charge they transfer to the ground; more than 90% of lightnings are negative and generally emerge from the bottom of a storm cloud, while the positive lightnings typically emerge from its upper levels or edges. Negative lightnings often occur in sequences, with the first (generally also the strongest) stroke and one or several subsequent strokes all proceeding along the same path, while positive lightnings usually consist of a single stroke. Occasionally, a sequence can also contain strokes of opposite polarity; such lightnings are sometimes classified separately as bipolar.

For first negative and positive strokes, the 50-th percentile (90-th percentile) values of the peak electric current are about 30 kA (80 kA) and 35 kA (250 kA), respectively, the median time from the onset of the current to its peak is about 5 µs and 20 µs, respectively, and the time from the peak of the current to its halving is about 75 and 230 µs, respectively (137–141). Following the IEC standard 60060-1 (142), the time course of the electric current of a lightning stroke can be approximated by a double exponential function

$$I(t) = I_\alpha \left( e^{-t/t_2} - e^{-t/t_1} \right) \tag{1a}$$

for which the best fit to the median parameter values for the electric current of the first negative stroke (30 kA peak value, 5 µs zero-to-peak time, and 75 µs halving time) is obtained with

$$I_\alpha = 31.76 \text{ kA}, \, t_1 = 1.077 \text{ µs}, \, t_2 = 106.6 \text{ µs} . \tag{1b}$$

An estimate of the electric field induced in an aqueous medium by $I(t)$ can be obtained by assuming that the medium is infinite and homogeneous, and hence the current flows through the medium radially from its point of entry. At a distance $r$ from this point, $I(t)$ is then distributed uniformly over a hemispherical surface with an area of $2\pi r^2$, resulting in a current density

$$J(r,t) = \frac{I(t)}{2\pi r^2} \tag{2}$$

which in a medium with electrical conductivity σ induces an electric field

$$E(r,t) = \frac{J(r,t)}{\sigma} = \frac{I(t)}{2\pi\sigma r^2} \quad . \tag{3}$$

The peak value of the induced electric field as a function of $r$ can then be estimated by inserting the peak value of $I(t)$

$$E_{max}(r) = \frac{\max(I(t))}{2\pi\sigma r^2} \quad . \tag{4}$$

On the microsecond time scale, heat flow due to temperature gradients can be disregarded, which allows the temperature increase caused by $I(t)$ to be estimated as

$$\Delta T(r,t) = \frac{1}{\rho c_p} \int_0^t E(r,\tau) J(r,\tau) d\tau = \frac{1}{\rho c_p} \int_0^t \frac{I^2(\tau)}{4\pi^2 \sigma r^4} d\tau = \frac{1}{4\pi^2 \sigma \rho c_p r^4} \int_0^t I^2(\tau) d\tau \tag{5}$$

where ρ and $c_p$ are the mass density and the specific heat capacity of the medium, respectively. The peak temperature increase as a function of $r$ can then be estimated by inserting the time course of $I(t)$ as given by Eq. 1 and evaluating the integral. The upper limit of integration can be set at some time after which $I(t)$ becomes negligible, but as $I^2(t)$ is integrable over the interval [0, +∞), this somewhat arbitrary choice can be avoided by writing

$$\Delta T_{max}(r) = \frac{1}{4\pi^2 \sigma \rho c_p r^4} \int_0^\infty I^2(\tau) d\tau \tag{6}$$

in which for aqueous media we can also set ρ ≈ 1 g/cm³ and $c_p$ ≈ 4.2 J/(g·K). For quantitative evaluation of Eqs. 4 and 6, the only remaining parameter is then the electrical conductivity of the medium, σ.

Natural aqueous environments exhibit a rather wide range of electrical conductivities. For rainwater, the values of σ range from 0.02 mS/cm (approximately double of the value for pure water) to 0.15 mS/cm (143,144); for river water, typical values are in the range from 0.1 mS/cm to 0.5 mS/cm (145,146); for lake water, they range from as low as 0.1 S/m for the lakes with the highest water throughput, up to 160 mS/cm for the most hypersaline of the salt lakes (147); for sea water, they are typically in the range from 20 to 60 mS/cm (148,149). Since both the induced electric field and the temperature increase are inversely proportional to σ, the radii of the regions of reversible poration, irreversible poration, and thermal damage (see Fig. 5) are therefore also strongly dependent on the type of aqueous environment exposed to a lightning. Table II gives the radial distances from the lightning stroke's point of entry characterized by various levels of temperature increase and induced electric field, as they result from Eqs. 4 and 6, for four electrical conductivities of the aqueous medium spanning from 0.1 to 100 mS/cm. The distances in the table correspond to the median parameter values for the electric current of the first negative lightning stroke, while for different peak values and/or durations of the current these distances are scaled accordingly.

Table II. The radial distances from the lightning's point of entry for various levels of temperature increase and induced electric field in the aqueous medium.

| effect | comment* | radial distance | | | |
|---|---|---|---|---|---|
| | | $\sigma = 0.1 \frac{mS}{cm}$ | $\sigma = 1 \frac{mS}{cm}$ | $\sigma = 10 \frac{mS}{cm}$ | $\sigma = 100 \frac{mS}{cm}$ |
| $\Delta T = 70°C$ | TD (150,151) | 14.6 cm | 8.19 cm | 4.60 cm | 2.59 cm |
| $\Delta T = 30°C$ | possible TD | 18.0 cm | 10.1 cm | 5.69 cm | 3.20 cm |
| $E = 30 \frac{kV}{cm}$ | IEP (60,62) | 39.9 cm | 12.6 cm | 3.99 cm | 1.26 cm |
| $E = 10 \frac{kV}{cm}$ | REP, EF, possible IEP | 69.1 cm | 21.9 cm | 6.91 cm | 2.19 cm |
| $E = 3 \frac{kV}{cm}$ | REP (60,62), EF (132–136) | 126 cm | 39.9 cm | 12.6 cm | 3.99 cm |
| $E = 1 \frac{kV}{cm}$ | possible REP and EF | 219 cm | 69.1 cm | 21.9 cm | 6.91 cm |

* TD: thermal damage; IEP/REP: irreversible / reversible electroporation;
EF: electrofusion (for organisms without the cell wall)

**Assessing the Feasibility of Lightning-triggered HGT.** The quantitative analysis presented above suggests that sufficiently close to the surface of natural aqueous environments with a sufficiently high density of prokaryotic population (i.e., a sufficiently high number of prokaryotic organisms per unit volume), lightning-triggered HGT may well be feasible; the ensuing question is then whether such densities do occur in environments exposed to lightnings. A comprehensive theoretical analysis of the dependence between the density of prokaryotic population and the probability of DNA released from one prokaryote by electroporation to come into contact with another prokaryote capable of transformation by this DNA (due to either reversible electroporation or natural competence) would contribute significantly to the understanding of this issue, particularly as it would allow to estimate the minimal population density required for non-negligible probability of lightning-triggered HGT. Still, the fact that transformation by means of natural competence occurs in a number of aquatic bacteria — see e.g. Table I in (30) — implies that at least in some aqueous environments, bacterial populations are sufficiently dense for the DNA released by some mechanism from one bacterium to come into contact with another bacterium and transform it. Lightning-triggered irreversible electroporation is one such mechanism of DNA release, and in regions with frequent thunderstorms it is perhaps also not insignificant compared to other natural causes of bacterial death.

Eqs. 4 and 6 imply, and Table II illustrates, that as electrical conductivity of the medium increases, both the region of lightning-induced DNA release and the region of lightning-induced ability of DNA uptake and transformation decrease in radius and thus in volume. Furthermore, Table II shows that in saline waters struck by a lightning, electric fields sufficient for irreversible electroporation generally also cause some extent of thermal DNA denaturation. This suggests that the likelihood of lightning-triggered HGT should be higher in freshwater than in saltwater environments; in particular, it should be the highest in freshwater habitats with the richest prokaryotic populations, such as open-air sewers.

Several experimental studies can also be viewed as providing tentative support for feasibility of lightning-triggered HGT among prokaryotes. In the early 1990s, as electroporation was becoming an established method of bacterial transformation, several researchers investigated the possibility of using electroporation also for DNA extraction, which would simplify the apparatus and streamline the protocol for electrotransformation (152–155). By exposing a mixture of donor and recipient bacteria to a single pulse, gene electrotransfer was obtained between two strains of *Escherichia coli* (152), from *Escherichia coli* into *Salmonella typhimurium* (153), and from *Escherichia coli* into *Pseudomonas aeruginosa* (154). The transformation efficiencies achieved in this manner were, however, by at least two to three orders of magnitude lower than if the donor bacteria were electroporated first, the supernatant subsequently isolated by centrifugation and transferred to the recipient bacteria, which were then electroporated separately (154). Furthermore, even the efficiencies achievable with the latter approach were by an order of magnitude lower than with DNA isolated by the standard procedure of extraction by lysis and purification in CsCl-ethidium bromide density gradients (154). The aim of these studies was to improve the methodology of bacterial transformation, and thus the approach based on electroporating a mixture of donor and recipient bacteria by a single pulse was largely dismissed as too inefficient for practical applications (154, 155), while the feasibility of its natural, lightning-triggered occurrence was only mentioned in one of these studies, by James Pfau and Philip Youderian, in their concluding sentence as a "speculation" (153).

The possible role of electroporation in bacterial evolution was contemplated again in 1995 by Jack Trevors in a review of advances in the understanding of molecular evolution in bacteria (31), where he mentioned electroporation as a method of bacterial transformation, and proceeded to state that the possibility of its natural occurrence "is an interesting concept" and that although "it may never be known if electrical discharges played a role in gene transfer during evolution, it is mentioned here to stimulate others to think about this process."

Six years later, when the importance of HGT in evolution was already much clearer, the subject of electroporation-triggered HGT was revived — this time with an explicit focus on lightnings as the source of electroporation and hence on natural feasibility — in two studies by the group of Pascal Simonet, in which they obtained electrotransfer of plasmid DNA into *Escherichia coli* (32) and two species of the genus *Pseudomonas* (33). Methodologically, these two studies did not differ considerably from many earlier investigations and applications of electrotransformation (see Table I), but they were apparently the first to attach importance to the similarity of the experimental conditions to those existing in nature, in particular by electroporating bacteria in soil (the native habitat of the investigated *Pseudomonas* species, although not of *E. coli*), and by attempting to make the exposure to electric pulses at least broadly comparable to lightning strokes. In the latter respect, it should be noted that the pulses were delivered through electrodes in direct contact with the soil sample, i.e. without an actual electric arc, and that in comparison to natural lightning strokes, the pulses were much lower in their peak current (6 A) and peak electric field induced in the sample (6.3 kV/cm), yet longer in their duration (time constant of 6 ms).

Several guidelines thus emerge for further experimental investigation of the feasibility of natural lightning-triggered HGT among prokaryotes:

- The exposure to the electric field in the laboratory environment should be as similar to those resulting from natural lightning strokes as reasonably achievable. Some degree of downscaling is difficult to avoid, but — as in the famous Miller-Urey experiments (156,157) — at least the physical principles governing lightnings should be emulated. Namely, in standard applications of electroporation, pulses are generated by a voltage source and delivered to electrodes in direct contact with the sample; modern generators maintain a preset voltage on the sample throughout the pulse, while simpler ones essentially discharge a capacitor through the sample, with the voltage decaying exponentially from a preset amplitude to zero. In contrast, a lightning stroke proceeds through a highly conductive channel (electric arc) created by electrical breakdown of the air separating the cloud and the ground, and the time course of the electric current and the voltage induced by it are neither rectangular nor purely exponentially decaying (see Eq. 1). Furthermore, in the ground the current does not flow towards a well-defined electrode, but dissipates roughly radially from its point of entry, and consequently the amplitude of the electric field it induces in the ground decreases rapidly with increasing distance from this point (see Fig. 5).

  While an electric current with a time course similar to those of lightning strokes could in principle be delivered to the sample by a programmable signal generator with a suitable voltage and current amplifier, a straightforward alternative is to avoid direct contact of one electrode with the sample and deliver the current through an actual electric arc generated in the air that separates the electrode from the sample. This yields a downscaled, but otherwise physically realistic emulation of a lightning stroke. The other electrode should be shaped and positioned in the sample as to ensure a roughly radial flow of the current from the point of its entry.

- The laboratory environment in which the studies are performed should be as free of conditions and substances absent from natural environments as possible; this should be the case throughout all the relevant stages of the experiment — at least from the start of the exposure to electric pulses to the evaluation of the resulting transformation and expression. Techniques such as centrifugation, nonnatural modifiers of osmolarity (e.g., sorbitol, mannitol) and enhancers of membrane fusibility (e.g., polyethylene glycol, fusogenic proteins) should therefore be avoided, albeit at a cost of decreased experimental resolution and/or increased apparatus complexity. Along the same lines, the chemical composition of the aqueous medium should resemble the composition of one of the naturally existing aqueous environments exposed to lightning strokes; for a proof of principle, appropriate concentrations of anorganic salts should suffice in this respect.

- The DNA molecules for which lightning-triggered HGT is being studied should be devoid of artificial modifications introduced to improve efficiency of transfer in general, and transferability among different species in particular. Successful lightning-triggered transformation with a shuttle vector or another type of engineered or altered DNA molecule does not necessarily imply that natural plasmid or chromosomal DNA can be transferred in the same manner. As with some of the guidelines above, this is likely to complicate the experiments, as it disqualifies some of the most widely used techniques of HGT assessment, in particular many varieties based on antibiotic resistance conferred by artificially modified plasmids.

Feasibility of natural lightning-triggered electrofusion should be studied along the same guidelines, with an additional requirement that no laboratory techniques of cell wall weakening or removal should be applied. As the existing data demonstrate rather conclusively that electrofusion of prokaryotes with an intact wall is not achievable, this restricts the experiments to the prokaryotes naturally lacking a cell wall, such as the species of the class Mollicutes, and potentially the naturally existing L-forms of various bacterial species.

In conclusion, based on the presented empirical evidence and theoretical considerations, it appears that electroporation is a plausible contributor to natural HGT among prokaryotes, especially in freshwater environments with rich prokaryotic populations. Much further work will likely be required to provide a reliable estimate of its current importance in comparison to conjugation, natural competence, and viral transduction. An upper bound for this estimate may be obtainable by studying lightning-triggered HGT in open-air sewers; a complementary line of investigation could perhaps be to try to attribute each of the identified occurrences of gene transfer among prokaryotes to one of the three biochemical HGT mechanisms, and see how many transfers – if any – remain unattributed. Still, as electroporation is a purely physical mechanism, and each of the three biochemical mechanisms must have developed during a certain stage of evolution, the relative importance of electroporation prior to these stages should have been correspondingly higher. Along similar lines, natural absence of the cell wall is rare in modern prokaryotes, but it is reasonable to assume that the wall has formed during a certain — albeit perhaps very early — stage of evolution, and at least prior to this, electrofusion could also have contributed importantly to natural HGT, yielding hybrid prokaryotes expressing the genetic material of their both precursors.

**ACKNOWLEDGMENTS.** This work was supported by the Slovenian Research Agency (Grant P2-0249). The research was conducted in the scope of the EBAM European Associated Laboratory (LEA) and within the networking efforts of the COST Action TD1104.